\documentclass[pdflatex,sn-mathphys-num, iicol]{sn-jnl}

\usepackage{graphicx}%
\usepackage{multirow}%
\usepackage{amsmath,amssymb,amsfonts}%
\usepackage{amsthm}%
\usepackage{mathrsfs}%
\usepackage[title]{appendix}%
\usepackage[dvipsnames]{xcolor}%
\usepackage{textcomp}%
\usepackage{manyfoot}%
\usepackage{booktabs}%
\usepackage{algorithm}%
\usepackage{algorithmicx}%
\usepackage{algpseudocode}%
\usepackage{listings}%
\usepackage{subcaption}
\usepackage{soul}
\usepackage[inline]{enumitem}%

\usepackage[capitalise]{cleveref}

\theoremstyle{thmstyleone}%
\theoremstyle{thmstyletwo}%

\theoremstyle{thmstylethree}%

\begin{document}

\title[PINNs for Electric Field Reconstruction in TPCs]{Physics-informed continuous normalizing flows to learn the electric field within a time-projection chamber}


\author*[1]{\fnm{Ivy} \sur{Li}}\email{il11@rice.edu}

\author[3,4]{\fnm{Peter} \sur{Gaemers}}

\author[1]{\fnm{Juehang} \sur{Qin}}

\author[1]{\fnm{Naija} \sur{Bruckner}}

\author[3,4]{\fnm{Maris} \sur{Arthurs}}

\author[3,4,5]{\fnm{Maria Elena} \sur{Monzani}}

\author[1,2]{\fnm{Christopher} \sur{Tunnell}}

\affil[1]{\orgdiv{Department of Physics and Astronomy}, \orgname{William Marsh Rice University}, \orgaddress{\city{Houston}, \postcode{77005}, \state{TX}, \country{United States}}}

\affil[2]{\orgdiv{Department of Computer Science}, \orgname{William Marsh Rice University}, \orgaddress{\city{Houston}, \postcode{77005}, \state{TX}, \country{United States}}}

\affil[3]{\orgdiv{SLAC National Accelerator Laboratory}, \orgname{Stanford University}, \orgaddress{\city{Menlo Park}, \postcode{94025}, \state{CA}, \country{United States}}}

\affil[4]{\orgdiv{Kavli Institute for Particle Astrophysics and Cosmology}, \orgname{Stanford University}, 
\orgaddress{
\city{Stanford}, \postcode{94305-4085}, \state{CA}, \country{United States}}}

\affil[5]{\orgname{Vatican Observatory}, \orgaddress{\city{Albano Laziale RM}, \postcode{0041}, \country{Italy}}}


\abstract{Accurate position reconstruction in noble-element time-projection chambers (TPCs) is critical for rare-event searches in astroparticle physics, yet is systematically limited by electric field distortions arising from charge accumulation on detector surfaces. Conventional data-driven field corrections suffer from three fundamental limitations: discretization artifacts that break smoothness and differentiability, lack of guaranteed consistency with Maxwell's equations, and statistical requirements of $\mathcal{O}(10^7)$ calibration events. We introduce a physics-informed continuous normalizing flow architecture that learns the electric field transformation directly from calibration data while enforcing the constraint of field conservativity through the model structure itself. Applied to simulated $^{83\mathrm{m}}$Kr calibration data in an XLZD-like dual-phase xenon TPC, our method achieves superior reconstruction accuracy compared to histogram-based corrections when trained on identical datasets, demonstrating viable performance with only $6\times10^5$ events---an order of magnitude reduction in calibration requirements. This approach enables practical monthly field monitoring campaigns, propagation of position uncertainties through differentiable transformations, and enhanced background discrimination in next-generation rare-event searches.}




\maketitle

\section{Introduction}\label{sec1}

Noble element time projection chambers are multipurpose observatories for astroparticle, particle, and nuclear physics; these detectors are used to search for dark matter~\cite{Planck:2018vyg, XENON:2023cxc, Bertone:2004pz, PandaX:2024qfu, DEAP-3600:2024szw, EXO-200:2022adi, LZ:2022lsv, DarkSide-50:2022qzh}, to probe for neutrinoless $\beta\beta$ decay~\cite{nEXO:2021ujk, Agostini:2022zub, LZ:2019qdm, XENON:2022evz}, and to study coherent elastic neutrino-nucleus scattering (CE$\nu$NS)~\cite{XENON:2024ijk, PandaX:2024muv, COHERENT:2017ipa}, and other rare events in fundamental physics~\cite{Aalbers:2022dzr}. The next generation of liquid noble element experiments are expected to be commissioned in the next 10 to 20 years, including but not limited to XLZD~\cite{Baudis:2024jnk}, DUNE~\cite{DUNE:2015lol}, PandaX-xT~\cite{PANDA-X:2024dlo}, and DarkSide-20k~\cite{DarkSide-20k:2017zyg}. The purpose of this paper is to develop computational methods for these future experiments.

Accurate parameter reconstruction presents a significant challenge for rare-event searches due to imperfect detector conditions~\cite{XENONCollaboration:2024bil}. For example, static charge builds up on dielectric surfaces inside TPCs to distort the electric field \cite{XENONnT:2023dvq}. As this charge builds up after the TPC has been sealed, direct \emph{in-situ} measurements of this distortion are unfeasible~\cite{XENON:2024wpa}. Consequently, reliable estimation of parameters that depend on the electric field, such as the positions of particle interactions, can be difficult.

We present the first application of physics-informed continuous normalizing flows to model electric fields in liquid noble-element TPCs, demonstrating a new approach that learns the electric field lines directly from data while incorporating Maxwell's equations as a physical constraint in the architecture. This offers three advantages over traditional field distortion correction (FDC) maps: the electric field model is constrained by Faraday's Law at the electrostatic limit, the smooth and differentiable transformation enables the propagation of position uncertainties, and this method reduces the amount of calibration data needed for detector commissioning. In this work, we focus on an XLZD-like detector~\cite{XLZD:2024nsu} but our method should be broadly applicable for single and dual-phase noble element TPCs.

\subsection{Noble-element time-projection chambers}

Liquefied xenon is a highly effective target for rare physics searches~\cite{Fujii:2015qqq, lansiart1976development, Aprile:2009dv, Chepel:2012sj, Heindl:2010zz}. When a particle scatters off a xenon atom, the stopping of the atom's nucleus or electron generates measurable signals. This detection method in liquid xenon TPCs has become an established particle detector technology in use since the 1970s. The working principle of dual-phase xenon TPCs is discussed in the book~\cite{aprile2006noble}, and details related to some specific realizations of the experiment are in~\cite{XENON:2024wpa, LZ:2019sgr, Li:2014rca}. Hardware support for electric field designs in particular is detailed in~\cite{XENONnT:2023dvq, Linehan:2021qnb}. 

In a dual-phase noble element TPC, the split liquid and gaseous volume enables separate measurements of ionization and scintillation signals, from which we can reconstruct a 3-dimensional interaction position. When a particle scatters off of a xenon atom inside the detector medium, this causes either an electron or the nucleus to recoil. The particle stopping deposits energy into the medium, thereby creating a prompt scintillation signal (S1) and releasing ionization electrons. These electrons drift due to the electric field toward the liquid-gas interface, and then are accelerated into the gas by a stronger extraction field, producing an amplified secondary ionization signal (S2). Both the S1 and S2 signals are detected by the large-area photosensor arrays~\cite{LopezParedes:2018kzu} at the top and bottom of the detector. This distribution of light detected by the photosensors is termed a \emph{hit pattern}.

Reconstructing the S2 position in the plane of the photosensors $(r_{\mathrm{S2}},\phi_{\mathrm{S2}})$ position of the S2 requires a sufficiently strong extraction field to ensure a highly localized S2 signal in the photosensor array~\cite{LUX:2017lif, LUX:2017bef, XENONCollaboration:2024bil}. Field-shaping rings are used to reduce electric field distortions and minimize the horizontal displacement of electrons~\cite{XENONnT:2023dvq}. The time delay between the S1 and S2 signals, or the drift time of the ionization electrons, provides a direct measure of the vertical depth of the interaction inside the TPC. With the drift information and reconstructed S2 signal position, we reconstruct the interaction vertex position $(r,\phi,z)$ in the liquid volume~\cite{XENONCollaboration:2024bil, LUX:2017bef}.

However, even with optimized electrode configurations and field-shaping systems, residual field distortions remain throughout the detector due to charge build-up on the polytetrafluoroethylene (PTFE) walls and geometric limitations~\cite{XENONnT:2023dvq}. Consequently, a fraction of the detector volume is partially charge-insensitive: ionization electrons are partially or completely lost due to electric field lines ending on the PTFE walls, and not every ionization electron from an interaction vertex is successfully extracted. We refer to this region as the \textit{charge insensitive volume}.

Furthermore, the field shaping ring that counteracts field leakage at the top of the detector from the anode must be biased to a higher voltage to prevent field leakage because the extraction field is stronger than the drift field. This draws more electrons near the top of the TPC away from the photosensors. These field distortions that cannot be eliminated by hardware design must be addressed through analysis techniques and data corrections~\cite{XENONCollaboration:2024bil}. 

Nonuniformities in the electric field distort the waveforms of S2 signals and shift the lateral position of the ionization electrons~\cite{LUX:2017mhm, XENONnT:2023dvq}. The reduction of field distortion greatly increases background rejection and the fiducial volume of correctly reconstructed events inside the TPC for analysis~\cite{XENONCollaboration:2024bil, LUX:2017bef}. The development of such field corrections in current generation experiments is covered in~\cref{sec:traditional-fdc} and our implementation of an FDC map to compare with the continuous normalizing flow model is covered in~\cref{sec:fdc-flow-compare}. 

\subsection{Field distortion correction $\Delta r$}\label{sec:traditional-fdc}

\begin{figure*}[htb]
    \centering
    \includegraphics[width=0.98\linewidth]{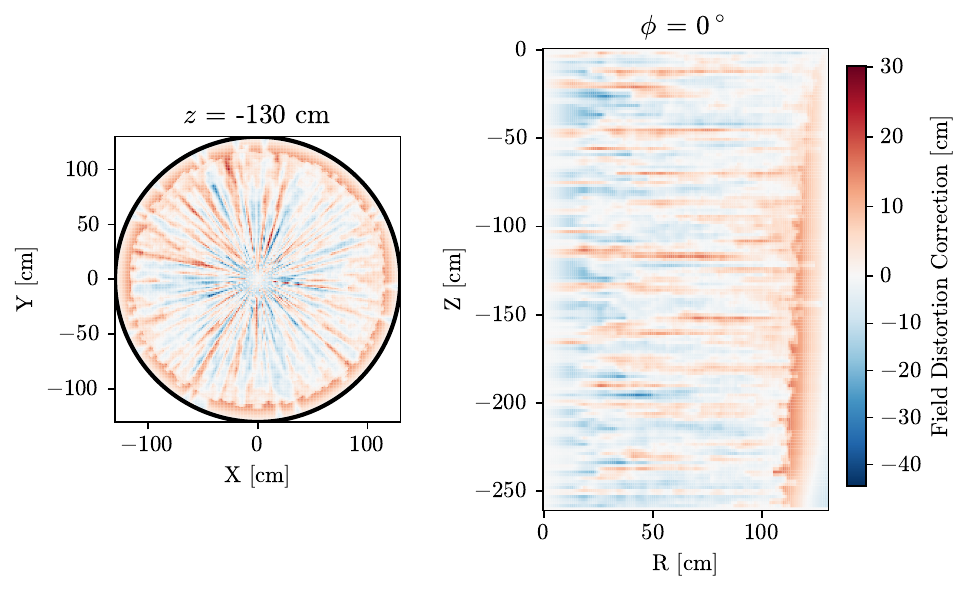}
\caption{Previous state-of-the-art: here we show the previous method for field distortion corrections. Our new method removes these unphysical artifacts in the depicted slices of the field distortion correction (FDC) map when correcting for field distortions. Slices in $z=-130$ cm and $\phi=0 ^{\circ}$ of the interpolated FDC map are shown following XENONnT's binning scheme of 95 $z$ bins, 100 $r^2$ bins, and 180 $\phi$ bins.}
\label{fig:tradfdc-slices}
\end{figure*}

In recent experiments, like LUX and XENONnT~\cite{LUX:2017bef, XENONnT:2023dvq}, field distortion correction maps are developed by calibrating the detector with a source that is uniformly distributed throughout the charge-sensitive volume such as ${}^{83\mathrm{m}}$Kr. As previously mentioned, the ionization electrons drift through the TPC following the electric field lines and the probability of detecting them corresponds with their interaction positions as based on the electron survival probability map. Starting from their interaction vertices, the ${}^{83\mathrm{m}}$Kr event positions shift radially inwards due to the charge accumulation on the PTFE walls which causes the electric field to funnel inwards. 

To recover the interaction vertices, a radial correction is developed for the final S2 positions measured at the top of the TPC. The distribution of reconstructed ${}^{83\mathrm{m}}$Kr event positions $(r_{\mathrm{S2}}, \phi_{\mathrm{S2}}, z)$ is compared with the distribution of expected interaction vertex positions $(r_{\mathrm{int}},\phi_{\mathrm{int}},z)$, which takes into account the charge insensitive volume. The detector volume is nearly cylindrical, so these distributions are binned over $(r, \phi, z)$, in which $r$ is the radial distance from the center, $\phi$ is the azimuthal angle, and $z$ is the depth. The correction is the radial difference between the corresponding percentiles of the reconstructed event position distribution and the expected position distribution. The field distortion correction map is defined for a given position
$(r, \phi, z) \rightarrow \Delta r$ in which the corrected interaction vertex position is
\begin{equation}\label{eq:fdcmap}
(r_{\mathrm{S2}}, \phi_{\mathrm{S2}}, z) \rightarrow (r_{\mathrm{S2}} + \Delta{r}, \phi_{\mathrm{S2}}, z).
\end{equation}
Current generation experiment FDC maps have several undesirable properties: 
\begin{enumerate*}
\item the correction map is not continuous and not guaranteed to be differentiable,
\item the transformation itself does not obey Maxwell's equations, and
\item learning this electric field transformation requires a large amount of data to have sufficient statistics spanning the entire detector volume
\end{enumerate*}.

Furthermore, due to the effects of dielectrics and detector geometry, modeling the electric field itself is challenging without finite-element methods which are computationally expensive~\cite{LUX:2017mhm, XENONnT:2023dvq}. This current FDC map approach cannot account for imperfections that develop during operations, such as displaced electrodes or uneven wire tensions that cannot be accurately measured until detector deconstruction. As the next generation of detectors comes online, the increase in detector volume may render this correction method unfeasible.

\section{Reconstructing electric fields using machine learning}\label{sec:e-field-ml}

We present a physics-informed continuous normalizing flow~\cite{NEURIPS2018_69386f6b, JMLR:v22:19-1028} to learn the electric field distortion from experimental data.~\cref{sec:node-intro} presents an overview how to implement continuous normalizing flows in modeling electric field lines.~\cref{sec:model-training} details our implementation for an XLZD-like detector, the continuous normalizing flow model architecture, and the training procedure.

Our method differs from traditional physics-informed neural networks~\cite{raissi2017physicsinformeddeeplearning} which typically incorporate physics constraints in the loss function. We instead enforce the physical constraint of a conservative electric field via model architecture as explained in~\cref{sec:node-intro} in which the model learns a scalar potential map instead of the electric field lines directly. We tested a more traditional physics informed model in which a curl term was added in the loss function, but we found that the method presented here performs similarly while not requiring tuning an additional hyperparameter that balances our likelihood loss with the physics-informed loss term. 

\subsection{Building physics-informed constraints into neural ODEs}\label{sec:node-intro}

We use a continuous normalizing flow to learn how electrons drift inside an XLZD-like TPC, following the papers on neural ODEs~\cite{NEURIPS2018_69386f6b} and the application of normalizing flows towards probabilistic inference~\cite{JMLR:v22:19-1028}. Neural ordinary differential equations (ODEs) are a class of machine learning models where an ODE solver is used to solve a differential equation,
\begin{equation}
    \frac{\partial\boldsymbol{r}(t)}{\partial t} = f_{\boldsymbol{\phi}}(\boldsymbol{r}(t),t),
\end{equation}
where $f_{\boldsymbol{\phi}}(\boldsymbol{r}(t),t)$ is a neural network with weights $\boldsymbol{\phi}$. This class of models transforms probability distributions, similar to normalizing flows, and are used as generative models (including as surrogate models for simulations and simulation-based inference); they are termed continuous normalizing flows when used in this capacity. In this work, we use such a neural ODE to model the trajectory of drifting electrons in a TPC.

Our approach differs from a direct application of continuous normalizing flows. Because the ODE we are modeling represents a physical process, we instead constrain the model such that the model only produces drift fields that represent physically realizable electric fields. Neglecting diffusion which is typically smaller than the mm-scale radial position resolution in these experiments~\cite{EXO-200:2016qyl, XENONCollaboration:2024bil, XENON:2024lbh}, we model the drifting of electrons in a TPC with
\begin{equation}\label{eq:ode}
    \frac{\partial \boldsymbol{r}(t)}{\partial t} =  -\mu(|\boldsymbol{E}|) \boldsymbol{E}(\boldsymbol{r}(t), t) \approx f_{\boldsymbol{\phi}}(\boldsymbol{r}(t),t),
\end{equation}
where $\boldsymbol{r}(t)$ is the position of a drifting electron as a function of time, the electric field is denoted $\boldsymbol{E}(\boldsymbol{r}(t),t)$, and the electron mobility is $\mu(|\boldsymbol{E}|)$.

Under electrostatic conditions, the electric field is a conservative field, such that
\begin{equation}
\begin{split}\label{eq:conservative_ode}
    \frac{\partial \boldsymbol{r}(t)}{\partial t} &=-\mu(|\boldsymbol{E}|) \boldsymbol{E}(\boldsymbol{r}(t))\\
    &= \mu(|\nabla V(\boldsymbol{r}(t), t)|) \nabla V(\boldsymbol{r}(t)).
\end{split}
\end{equation}

We simplify the model because the drift field is predominantly in the vertical direction that we define as the $z$-direction and that the drift velocity and electric field magnitude is largely constant (varying by $\sim 0.1\%$ in our simulation). This approximation is used in the FDC map correction procedure for XENONnT~\cite{XENONCollaboration:2024bil}. With this approximation, we decompose Eq.~\eqref{eq:ode} into
\begin{equation}\label{eq:drift_v_approx}
\begin{split}
    z(t) &= z(0)+v_\mathrm{d} t\\
    \frac{\partial \boldsymbol{s}(t)}{\partial t} &=  -\mu_\mathrm{avg} \boldsymbol{E_s}(\boldsymbol{s}(t), z)\\
    &= -\mu_\mathrm{avg} \boldsymbol{E_s}(\boldsymbol{s}(t), z_0+v_\mathrm{d} t)
\end{split}
\end{equation}
where 
\begin{equation}
    \boldsymbol{r}(t) = 
    \begin{pmatrix}
        \boldsymbol{s}(t)\\
        z(t)
    \end{pmatrix},
\end{equation}
$\mu_\mathrm{avg}$ is the average electron mobility, and $\boldsymbol{E_s}(\boldsymbol{s}(t), z(t))$ denotes the electric field in the transverse ($\boldsymbol{s}$) direction. We remove the z-dependence of the $\boldsymbol{s}(t)$ differential equation by defining the top of the TPC as $z=0$ and the time at which the electron arrives at the top of the TPC as $t_\mathrm{max}=0$, and thus defining the initial time at which the ionization signal is produced as $t=z_\mathrm{true}/v_\mathrm{d}$, because this definition gives us $z(0) = 0$ and $z(t)=v_\mathrm{d}t$. Note that the initial time and the true $z$-position are both negative in our coordinate definition.

The neural ODE representing this simplified problem is then
\begin{equation}\label{eq:2d_neural_ode_f}
\begin{split}
    \frac{\partial \boldsymbol{s}(t)}{\partial t} &= -\mu_\mathrm{avg} \boldsymbol{E_s}(\boldsymbol{s}(t), v_\mathrm{d} t)\\
    &\approx \boldsymbol{f'}_{\boldsymbol{\phi}}(\boldsymbol{s}(t), t)
\end{split}
\end{equation}
where $\boldsymbol{f'}_{\boldsymbol{\phi}}(\boldsymbol{s}(t), t)$ represents the two-dimensional neural ODE function in the transverse direction.

To ensure that our neural ODE is constrained to models that satisfy the electrostatic constraint in~\cref{eq:conservative_ode}, we define
\begin{equation}
    \boldsymbol{f'}_{\boldsymbol{\phi}}(\boldsymbol{s}(t), t) = -\nabla_{\boldsymbol{s}} g_{\boldsymbol{\phi}}(\boldsymbol{s}(t), t),
\end{equation}
and instead solve
\begin{equation}
    \frac{\partial \boldsymbol{s}(t)}{\partial t} = -\nabla_{\boldsymbol{s}} g_{\boldsymbol{\phi}}(\boldsymbol{s}(t),t).
\end{equation}

\begin{figure}[htb]
    \centering
    \includegraphics[width=\linewidth]{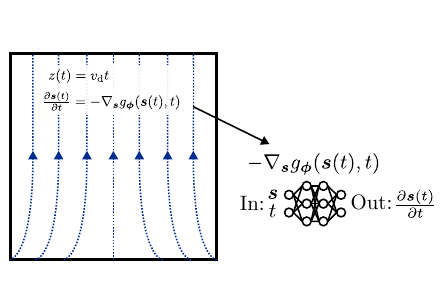}
    \caption{Diagram of continuous normalizing flow model. We represent a point within the box (TPC) with its transverse position $\boldsymbol{s}$ and its time $t$ which corresponds to depth $z(t)=v_{d} t$ such that $v_d$ is the field drift velocity. The neural network takes point as its input and returns how that position transforms over time, as governed by the neural network which represents the negative transverse gradient of a function $g_{\phi}$ that corresponds to an approximation of a scaled scalar potential.}
    \label{fig:cnf_architecture}
\end{figure}

The physical interpretation of $g_{\boldsymbol{\phi}}(\boldsymbol{s}(t), t)$ is that it is an approximation of the scaled scalar potential for every $z$-slice; that is, while the $z$-derivative of $g_{\boldsymbol{\phi}}$ does not correspond to the electric field in $z$, the $\boldsymbol{s}$-derivative represents the inverse of the electric potential times the electron mobility up to the constant term such that
\begin{equation}
    \nabla_{\boldsymbol{s}} g_{\boldsymbol{\phi}}(\boldsymbol{s}(t), t) \approx -\mu_\mathrm{avg}\nabla_{\boldsymbol{s}}V\left(\boldsymbol{r}(t)\right).
\end{equation}
Given $g_{\boldsymbol{\phi}}(\boldsymbol{s}, z/v_\mathrm{d})$, we can thus reconstruct the scalar potential as
\begin{equation}\label{eq:reconstruct_V_from_g}
\begin{split}
    V(\boldsymbol{s}, z) =& \frac{zV_\mathrm{cathode}}{z_\mathrm{cathode}} - \\
    & \quad \frac{g_{\boldsymbol{\phi}}(\boldsymbol{s}, z/v_\mathrm{d})-g_{\boldsymbol{\phi}}((0,0), z/v_\mathrm{d})}{\mu_\mathrm{avg}} ,
\end{split}
\end{equation}
where $V_\mathrm{cathode}$ is the voltage at the bottom electrode of the TPC, $z_\mathrm{cathode}$ is the $z$-position of the cathode, and the voltage at the top gate is taken to be $0 \, \mathrm{V}$.

A schematic of our model is shown in~\cref{fig:cnf_architecture}.

To compute the position at the top of the TPC given an initial true position $(\boldsymbol{s}_\mathrm{true}, z_\mathrm{true})$, then, we solve Eq.~\eqref{eq:drift_v_approx}, giving us
\begin{equation}
\begin{split}
    t_\mathrm{d} &= -z_\mathrm{true}/v_\mathrm{d}\\
    z(0) &= z_\mathrm{top} = 0,\\
    \boldsymbol{s}(-t_\mathrm{d}) &= \boldsymbol{s}_\mathrm{true}\\
    \boldsymbol{s}(0) &= \boldsymbol{s}_\mathrm{true} - \int^{0}_{-t_\mathrm{d}}  \nabla_{\boldsymbol{s}} g_{\boldsymbol{\phi}}(\boldsymbol{s}(\tau), \tau) d\tau.
\end{split}
\end{equation}

When training, we integrate backwards in time for the measured drift time to find the initial values given the observed S2 position $(\boldsymbol{s}(0), z(0))$ and drift time $t_d$:
\begin{equation}\label{eq:inverse_integral}
\begin{split}
    z(-t_\mathrm{d}) &= -v_\mathrm{d} t_\mathrm{d},\\
    \boldsymbol{s}(-t_d) &= \boldsymbol{s}(0) - \int^{-t_\mathrm{d}}_{0}  \nabla_{\boldsymbol{s}} g_{\boldsymbol{\phi}}(\boldsymbol{s}(\tau), \tau) d\tau.
\end{split}
\end{equation}

In practice, the inverse integral in~\cref{eq:inverse_integral} that transforms observed positions into true positions is defined as the ``forward transformation" of the continuous normalizing flow in our codebase, as it is the direction that is used in both training and when deployed to correct position data, however this is a codebase convention that is mathematically equivalent with the presentation in this paper.

\subsection{Model architecture and training}\label{sec:model-training}

Training our model involves solving~\cref{eq:drift_v_approx} for multiple events, and ensuring the distribution of those events corresponds to the ideal distribution, which is one that is uniform throughout most of the detector, but which differs near the edges due to charge loss. The loss of events near the edges of the detector is modeled via the simulated electron survival probability as is done for the histogram-based FDC map (see~\cref{sec:traditional-fdc}).

To evaluate how well our method would work on a real experiment, we use simulated $^{83\mathrm{m}}$Kr events as done in~\cref{sec:traditional-fdc}. While we have access to the true positions for simulated data, this would not be accessible when trying to use our method on real data. As such, we cannot simply train our neural ODE to minimize the reconstruction error, for example, by using the mean-squared error between the reconstructed position and the true position as the loss function. Instead, we need to use a training objective that targets the same property of $^{83\mathrm{m}}$Kr calibrations that enable histogram-based field distortion correction--the uniformity of the true positions within the charge insensitive volume.

Earlier, in~\cref{sec:node-intro}, we have framed our method as a neural ODE that we use to approximate the equations of motion of electrons drifting in a TPC; in this section, however, we will interpret the same method as a continuous normalizing flow that transforms probability densities as our training amounts to solving a simulation-based inference problem~\cite{Cranmer2019TheFO}. This approach of training a neural ODE as a density estimator is key to enabling the use of $^{83\mathrm{m}}$Kr calibration data without access to the ground truth positions.

As this is a density estimation problem similar to neural posterior estimation, we use the negative log likelihood (NLL) as the loss function to train our continuous normalizing flow~\cite{zeghal2023neural}. In doing so, we optimize our neural network to find a transformation from the observed distribution of $^{83\mathrm{m}}$Kr events to the expected distribution of $^{83m}$Kr events, while obeying the physics-informed constraints given in~\cref{sec:e-field-ml}. The likelihood of $^{83\mathrm{m}}$Kr events is modeled using the following approximation of the electron survival probability termed $p_\mathrm{surv}(\boldsymbol{r})$. $p_\mathrm{surv}(\boldsymbol{r})$ is modeled as unity within the volume of the TPC without charge loss. As we move towards the edge of the TPC towards the charge insensitive volume, $p_\mathrm{surv}(\boldsymbol{r})$ decreases; however, we limit the minimum value to a set value $10^{-3}$ instead of letting it reach zero. This is then followed by an exponential dropoff with increasing radius with $\lambda=100\,\mathrm{cm}$. The reason the electron survival probability is made to roll off towards $10^{-3}$ instead of zero, followed by an exponential dropoff, is to ensure that the log-likelihood for events constructed in the charge-insensitive volume or outside of the TPC is finite and produces useful gradients for neural network training.

To emulate what would happen during a real detector calibration while using simulated events, we only give ourselves access to the hit pattern of each event and the drift time for $^{83\mathrm{m}}$Kr events simulated following~\cref{sec:simulation}. The $z$-position is reconstructed from drift time by multiplying it with the drift velocity, while the observed S2 position in $x,y$ is reconstructed using the probabilistic position reconstruction method given in~\cref{sec:probabilistic-posrec}. This probabilistic position reconstruction gives us a distribution of positions as opposed to a single position. The likelihood of the $i^\mathrm{th}$ event marginalized over possible S2 positions is thus 
\begin{equation}\label{eq:single_likelihood_integral}
\begin{split}
\ell^i(\boldsymbol{\phi}) 
=& \int \, \Big[p_{\mathrm{pos}}^i(\boldsymbol{s}_\mathrm{S2})\cdot\\
&\qquad p'_{\text{surv}}(\boldsymbol{s}_\mathrm{S2}, z_\mathrm{reconst},  \boldsymbol{\phi})\Big]d\boldsymbol{s}_\mathrm{S2}, \\
p'_{\text{surv}}&(\boldsymbol{s}_\mathrm{S2}, z_\mathrm{reconst},  \boldsymbol{\phi}) =\\
& p_{\text{surv}}(\boldsymbol{h}(\boldsymbol{s}_\mathrm{S2}, z_\mathrm{reconst},  \boldsymbol{\phi}))\left|\frac{\partial \boldsymbol{h}}{\partial \boldsymbol{s}_\mathrm{S2}}\right|, 
\end{split}
\end{equation}

where $p_{\mathrm{pos}}^i(\boldsymbol{s}_\mathrm{S2})$ is the distribution of S2 positions, and $\boldsymbol{h}(\boldsymbol{s}_\mathrm{S2}, z_\mathrm{reconst}, \boldsymbol{\phi})$ is the transformation between uncorrected S2 positions and corrected positions as represented by the neural ODE with weights $\boldsymbol{\phi}$. The determinant of the Jacobian of the transformation $\left|{\partial \boldsymbol{h}}/{\partial \boldsymbol{s}_\mathrm{S2}}\right|$ appears following the rules for transformation of probability densities~\cite{Amaral:2024edw}. As our transformation comes from solving a neural ODE, it is evaluated as the integral of the trace of the Jacobian~\cite{NEURIPS2018_69386f6b, grathwohl2018ffjord} (see~\cref{sec:trace-jac-appendix})
\begin{equation}\label{eq:jac-trace}
    \left|\frac{\partial \boldsymbol{h}}{\partial \boldsymbol{s}_\mathrm{S2}}\right| = \int_0^{-t_d} \mathrm{Tr}\left[\frac{\partial \boldsymbol{f}'_{\boldsymbol{\phi}}}{\partial \boldsymbol{s}(\tau)}\right] d\tau.
\end{equation}

As we can sample from $p_{\mathrm{pos}}^i(\boldsymbol{s}_\mathrm{S2})$, \cref{eq:single_likelihood_integral} can be approximated using Monte Carlo integration, giving us
\begin{equation}
\begin{split}
    \ell^i(\boldsymbol{\phi}) \approx & \frac{1}{N}\sum^N_j p_{\text{surv}}(\boldsymbol{h}(\boldsymbol{s}_\mathrm{S2}^{ij}, z_\mathrm{reconst}, \boldsymbol{\phi}))\left|\frac{\partial \boldsymbol{h}}{\partial \boldsymbol{s}_\mathrm{S2}}\right|,\\
&\mathrm{where}\quad\boldsymbol{s}_\mathrm{S2}^{ij} \sim p_{\mathrm{pos}}^i(\boldsymbol{s}_\mathrm{S2}).
\end{split}
\end{equation}

We can then compute the negative log-likelihood loss function across $M$ data points as
\begin{equation}
    \mathcal{L}(\boldsymbol{\phi}) = -\frac{1}{M}\sum^M_i \log \ell^i(\boldsymbol{\phi}).
\end{equation}

Further training details and model hyperparameters can be found in~\cref{sec:training-appendix}.

\section{Simulation Framework}\label{sec:simulation}

To train and validate our physics-informed continuous normalizing flow, we require a realistic simulation of detector response with known electric field distortions. We develop a next-generation liquid xenon TPC simulation based on XLZD design parameters within the \texttt{fuse}~\cite{henning_schulze_eissing_2025_15782421} and \texttt{strax}~\cite{jelle_aalbers_2024_11355772} frameworks, which are existing open source software packages for the XENONnT simulation chain and the data analysis pipeline for general xenon TPCs respectively. This simulation generates training data where both the true electric field and the true interaction positions are known, enabling supervised learning of the field distortion corrections.

\subsection{Detector Geometry and Electric Field Calculation}\label{sec:simulator-efield-calculation}

Our simulated detector uses XLZD design parameters: a cylindrical TPC with height $h_{\mathrm{TPC}} = 259.92$ cm and radius $R_{\mathrm{TPC}} = 129.96$ cm. We employ the same field cage design as XENONnT~\cite{XENONnT:2023dvq}, scaled to XLZD dimensions to maintain realistic field-shaping ring configurations.

\subsubsection{Electric Field Computation with FEniCS}

We calculate the electric potential $\phi(r,z)$ by solving Laplace's equation in cylindrical coordinates using the open-source software, FEniCS (DOLFINx)~\cite{BarattaEtal2023, ScroggsEtal2022}:
\begin{equation}
\nabla^2 \phi = 0.
\end{equation}
Assuming cylindrical symmetry, we solve on a 2D mesh in $(r,z)$ with the following boundary conditions:
\begin{itemize}
    \item \textbf{Gate} (top, $z = 0$): Dirichlet condition $\phi = 0$ V
    \item \textbf{Cathode} (bottom, $z = -h_{\mathrm{TPC}}$): Dirichlet condition $\phi = -100$ kV  
    \item \textbf{PTFE wall} ($r = R_{\mathrm{TPC}}$): Mixed boundary condition accounting for surface charge accumulation
\end{itemize}
The electric field is recovered as $\mathbf{E} = -\nabla \phi$.

\subsubsection{Wall Charge Accumulation Model}

Electric field distortions arise primarily from charge buildup on the insulating PTFE walls during detector operation. Following measurements from XENONnT~\cite{XENONnT:2023dvq} and LUX~\cite{LUX:2017mhm}, we implement a linear surface charge density model:
\begin{equation}
\sigma_w(z) = \lambda \cdot \frac{z}{h_{\mathrm{TPC}}} + \sigma_{\mathrm{top}},
\end{equation}
where $\sigma_{\mathrm{top}}$ is the surface charge density at the gate and $\lambda$ parameterizes the linear gradient. We scaled the XENONnT charge-up parameters by the ratio of TPC heights to obtain realistic values for XLZD. The surface charge density is incorporated into the FEniCS calculation through a mixed boundary condition on the PTFE surface that relates the normal component of the electric field to $\sigma_w$.

\subsection{Electron Drift Simulation}

From the computed electric field, we derive three critical maps that govern electron transport from interaction vertices to the liquid-gas interface.

\subsubsection{Drift Field Map}

For each point on a fine mesh grid in $(r, z)$, we integrate the electron trajectory along the electric field lines to determine the final position at the liquid-gas interface. The integration follows:
\begin{equation}
\frac{d\mathbf{r}}{dt} = \mathbf{v}_d(\mathbf{E}(\mathbf{r})),
\end{equation}
where the drift velocity $\mathbf{v}_d$ depends on the local electric field strength according to empirical field-velocity relationships from~\cite{Hogenbirk_2018}. This produces a mapping $(r_{\mathrm{init}}, z_{\mathrm{init}}) \rightarrow (r_{\mathrm{final}}, z_{\mathrm{final}} = 0$) that quantifies the radial displacement caused by field inhomogeneities. Near the walls, electrons experience radial displacements of several centimeters, directly impacting reconstructed S2 positions.

\begin{figure}[!htbp]
    \includegraphics[width=0.95\linewidth]{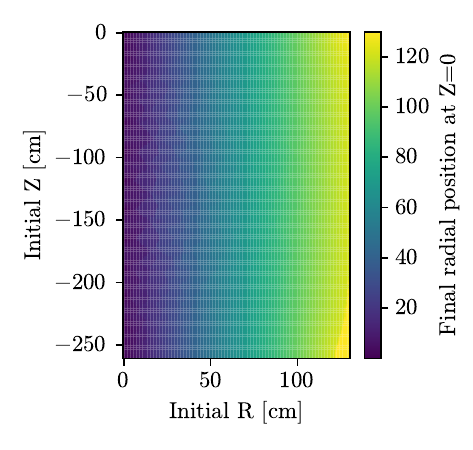}
    \caption{Final radial position at the liquid-gas interface $z = 0$ cm as a function of initial radius and depth, showing the distortion caused by nonuniform electric fields.}
    \label{fig:sim_drift_final_position}
\end{figure}

To achieve higher spatial resolution near the TPC boundaries, where field gradients are steepest, we used an adaptive mesh with increased granularity at the edges. The integration is performed using Euler's method with time step $\Delta t = 10^{-6}$ s and continues until the electron reaches the liquid-gas interface or is lost to the wall. The resulting map is shown in  \cref{fig:sim_drift_final_position}.

\subsubsection{Drift Velocity Map}

During trajectory integration, we calculate the local electric field strength $|\mathbf{E}(\mathbf{r}(t))|$ along each path and convert it into the instantaneous drift velocity using the empirical relationships from~\cite{Hogenbirk_2018}. We compute the path-averaged drift velocity:
\begin{equation}
\langle v_d \rangle = \frac{1}{t_{\mathrm{drift}}} \int_0^{t_{\mathrm{drift}}} v_d(|\mathbf{E}(\mathbf{r}(t))|) \, dt.
\end{equation}
This map enables accurate drift time calculations needed for electron lifetime corrections and vertical position reconstruction.

\subsubsection{Electron Survival Probability Map}

\begin{figure}[htbp]
    \begin{subfigure}[b]{0.48\textwidth}
        \centering
        \includegraphics[width=0.95\linewidth]{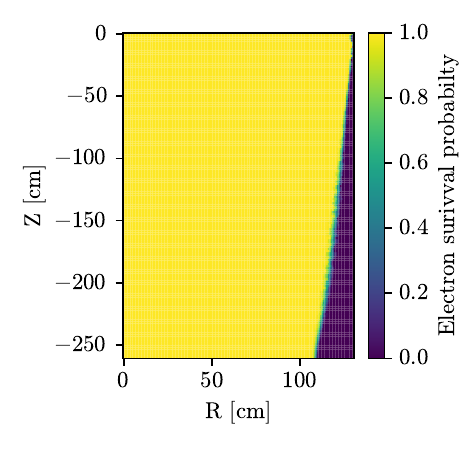}
        \label{fig:second_image}
    \end{subfigure}
    \caption{Electron survival probability map, accounting for radial diffusion and loss to the detector walls. Zero-probability regions define the charge-insensitive volume.
            }
    \label{fig:sim_electron_survival_probability}
\end{figure}

Electrons are lost during drift because diffusion has carried them into the PTFE walls. We model transverse diffusion as a Gaussian process with the diffusion coefficient $D_T = 55$ cm$^2$/s~\cite{albert2017measurement}. For an electron cloud that originates at position $(r, z)$, we track both the centroid position and the cloud width:
\begin{equation}
\sigma_{\mathrm{cloud}}^2(t) = 2 D_T t.
\end{equation}
At each integration step, we calculate the fraction of the Gaussian electron cloud with $r < R_{\mathrm{TPC}}$. The survival probability is the product of these fractions over the entire drift path. Regions with zero survival probability define the charge-insensitive volume of the detector. This modeling ensures a smooth survival probability distribution without discontinuities, which is essential for training the continuous normalizing flow. The survival probability map is shown in \cref{fig:sim_electron_survival_probability}.

Note that while we model diffusion to calculate survival probability, we do not track the final cloud width, as this is not needed for the S2 pattern generation described below.

\subsection{Optical Response Maps}

The optical response of the detector is characterized through Monte Carlo photon transport simulations using the XLZD Geant4 implementation in BACCARAT~\cite{Akerib_2021_sim}. We simulate $N_{\gamma} =  10^7 $ photons to construct two pattern maps.

For primary scintillation (S1), photons are generated uniformly throughout the active TPC volume. The map records the probability that a photon generated at position $(x, y, z)$ reaches the photocathode of each photomultiplier tube (PMT). The map is binned in [X $\times$ Y $\times$ Z] voxels and is not normalized, as approximately 50\% of the generated photons are absorbed by the detector materials before reaching a PMT. To account for quantum efficiency, all map values are multiplied by a factor of 0.3, representing the 30\% detection probability at the PMT photocathode~\cite{lung2012characterization}.

For secondary scintillation (S2), photon generation is restricted to the gas gap region between the gate and the top of the TPC, representing electroluminescence from extracted electrons. The map is binned in [X $\times$ Y] and normalized such that $\sum_{\mathrm{PMT}} P_{\mathrm{PMT}}(x,y) = 1$ for each position. 

For both maps, position-dependent light collection is interpolated using weighted nearest neighbors when sampling during event simulation.

\subsection{Event Simulation Workflow}
The user provides initial interaction parameters: interaction vertex positions $(x, y, z)$, interaction types (electronic recoil, nuclear recoil), energies, and times. Given these initial interaction parameters, events are simulated as described below.

\paragraph{Light and Charge Yield Generation}
For each interaction, we evaluate the local electric field strength at the vertex position from the electric field map. Using this field strength, the deposited energy, and the interaction type, we sample from NEST~\cite{jason_brodsky_2019_2535713} to determine the number of scintillation photons ($n_{\gamma}^{\mathrm{S1}}$) and ionization electrons ($n_e$).

\paragraph{S1 Signal Generation}
The $n_{\gamma}^{\mathrm{S1}}$ primary scintillation photons are distributed among the PMTs according to the S1 pattern map at the interaction position $(x, y, z)$. Each photon generates one photoelectron upon detection.

\paragraph{Electron Drift and Loss}
Electrons drift toward the liquid-gas interface following the drift field map. A fraction of electrons is lost due to attachment to impurities during drift. We model this using an electron lifetime $\tau_e = 10$ ms (a realistic, slightly conservative value for XLZD) and the drift time calculated from the velocity map:
\begin{equation}
n_e^{\mathrm{surviving}} = n_e \cdot \exp\left(-\frac{t_{\mathrm{drift}}}{\tau_e}\right).
\end{equation}
The surviving electrons are propagated to position $(x_{\mathrm{final}}, y_{\mathrm{final}}, z = 0 $ using the drift field map. The electron survival probability map (accounting for diffusion losses) is then applied:
\begin{equation}
n_e^{\mathrm{extracted}} = n_e^{\mathrm{surviving}} \cdot P_{\mathrm{survival}}(x, y, z).
\end{equation}

\paragraph{S2 Signal Generation}
The extracted electrons produce secondary scintillation in the gas gap. We assume a secondary scintillation gain of $g_2 = 60$ photons/electron (consistent with LZ measurements and conservative for XLZD). The total number of S2 photons is:
\begin{equation}
n_{\gamma}^{\mathrm{S2}} = n_e^{\mathrm{extracted}} \cdot g_2.
\end{equation}
These photons are distributed among the top PMT array according to the S2 pattern map at position $(x_{\mathrm{final}}, y_{\mathrm{final}})$. As with S1, each detected photon produces one photoelectron.

\paragraph{Simplifications}
To maintain computational efficiency while preserving the essential physics, we make two simplifications: (1) we assume each detected photon produces exactly one photoelectron (no PMT gain fluctuations or electronic noise), and (2) we do not track electron cloud sizes beyond calculating survival probability. For large signals relevant to electric field calibration with $^{83\mathrm{m}}$Kr, PMT noise is negligible and these simplifications do not affect the quality of position reconstruction or field learning.

\begin{figure}[htbp]
    \centering
    \includegraphics[width=\linewidth]{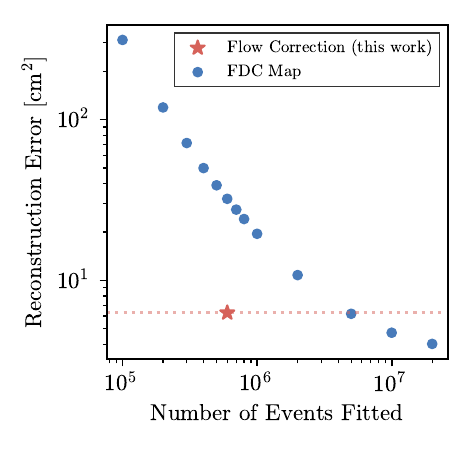}
    \caption{The continuous normalizing flow achieves a comparable reconstruction error with approximately an order of magnitude less calibration data than a traditional field distortion correction (FDC) map. The flow shown in this figure was trained on $6 \times 10^{5}$ events and it performs comparably with a field distortion map that was fit on $5 \times 10^6$ events. The reconstruction error displayed in this figure is the mean squared error on an independent test dataset of $5 \times 10^{5}$ events, comparing corrected positions with their corresponding ground truth interaction vertices.}
    \label{fig:mse-comparison}
\end{figure}

\begin{figure*}[htbp]
    \centering
    \includegraphics[width=\linewidth]{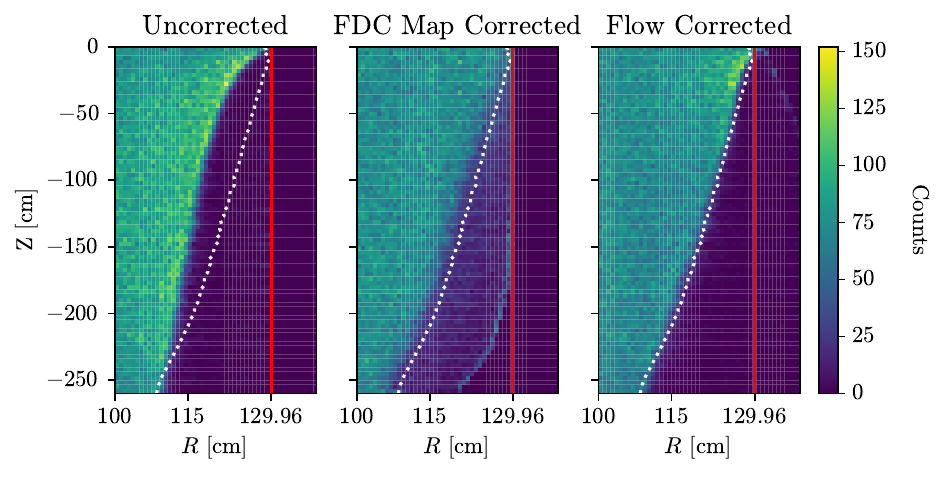}
    \caption{The figures illustrate the distribution of $(R,Z)$ where R is the radius and $Z$ is the depth of the interaction before and after a transverse correction has been applied. The red line marks the edge of the TPC at $r=129.96 \,\text{cm}$ and the white dotted line marks the edge of the charge insensitive volume, defined by the $p_\mathrm{surv}=0.5$ contour. The left figure shows the distribution of positions as inferred from the probabilistic position reconstruction model before corrections. The center figure shows the corrected distribution after applying the traditional field distortion correction (FDC) map as shown in~\cref{fig:tradfdc-slices}. The right figure shows the corrected distribution after the continuous normalizing flow transformation. The distribution is binned over $R^2$ instead of $R$ to ensure equal area exposure at different radii of a circle.}
    \label{fig:distribution-compare}
\end{figure*}

\begin{figure*}[htb]
    \centering
    \includegraphics[width=0.9\linewidth]{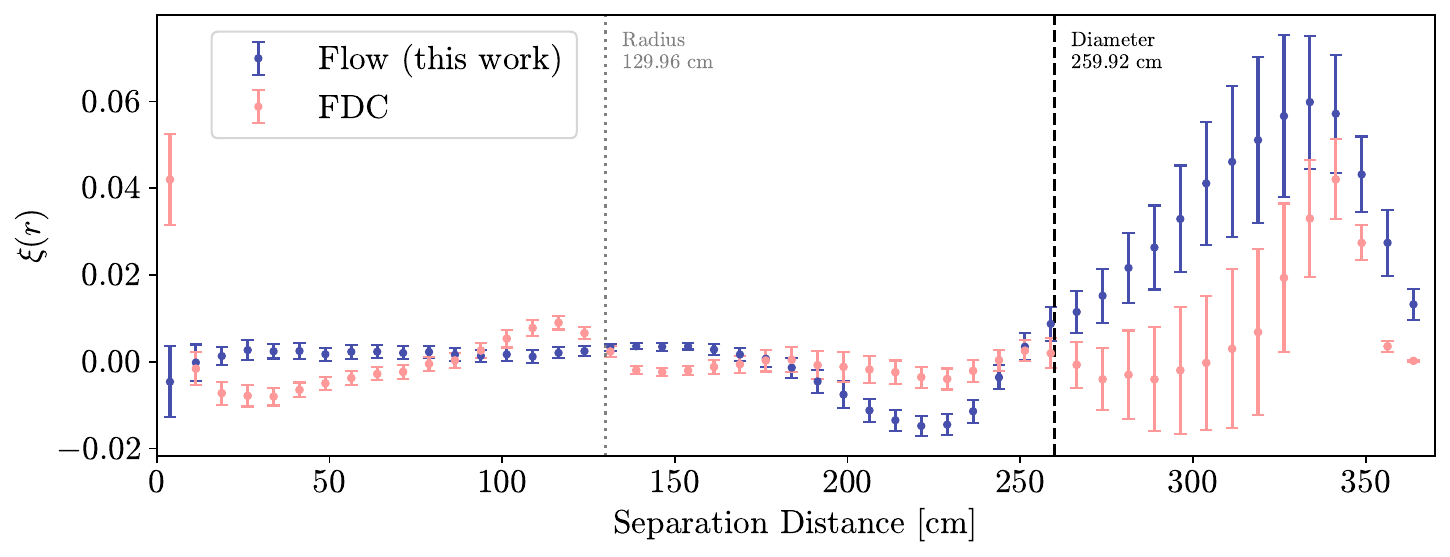}
    \caption{The flow preserves more local pairwise distances and clustering under one detector radius of separation distance than the field distortion correction (FDC) map. The figure compares the normalized point correlation functions of the flow corrected positions and FDC map corrected positions from the same test dataset of $5 \times 10^5$ events, and error bars were calculated from bootstrapping, or resampling $5 \times 10^4$ events from the dataset. These correlation functions have been normalized by subtracting the ground truth correlation function, and as such, overdensities are values above 0 and underdensities are below 0 at a given separation distance.}
    \label{fig:bootstrapped-2pcf}
\end{figure*}

\section{Results}

\subsection{Generating a field distortion correction map}
To establish a baseline, we generate a histogram-based FDC map from simulated ${}^{83\mathrm{m}}$Kr calibration events,  following the binning scheme of XENONnT~\cite{XENONCollaboration:2024bil}. To study the reconstruction error of the FDC map, we simulate varying numbers of ${}^{83\mathrm{m}}$Kr events as shown in~\cref{fig:mse-comparison}. Following the same procedure as the flow training, for each event, we reconstruct the S2 positions and depths, and then select successfully extracted events with extraction position reconstruction for the FDC map. When not specified, FDC in the rest of this paper will refer to the histogram-based FDC map that was fit to the same $6 \times 10^5$ event dataset that the flow was trained on.

To generate the map, the distribution of these selected reconstructed S2 positions and reconstructed $z$ (the product of the event's drift time and the average drift velocity from simulation) is binned over 95 slices in $z$ and 180 bins in $\phi$. For each $z$ and $\phi$ bin, 100 bins of $r^2$ are compared to compute the difference in $r$ between the corresponding percentiles of the reconstructed position distribution and the survival probability map distribution. We apply a linear interpolator to this map to obtain a field distortion correction for any given $(r_{\mathrm{S2}}, \phi_{\mathrm{S2}}, z_{\mathrm{reconst}})$. One $z$ slice and $\phi$ slice of the FDC map are shown in~\cref{fig:tradfdc-slices} which show discontinuities and artifacts from the binning scheme similar to the correction maps from XENONnT~\cite{XENONCollaboration:2024bil} and LUX~\cite{LUX:2017bef}.

\subsection{Comparing flow performance with field distortion correction map}\label{sec:fdc-flow-compare}

\begin{figure*}[htbp]
    \centering
    \includegraphics[width=0.8\linewidth]{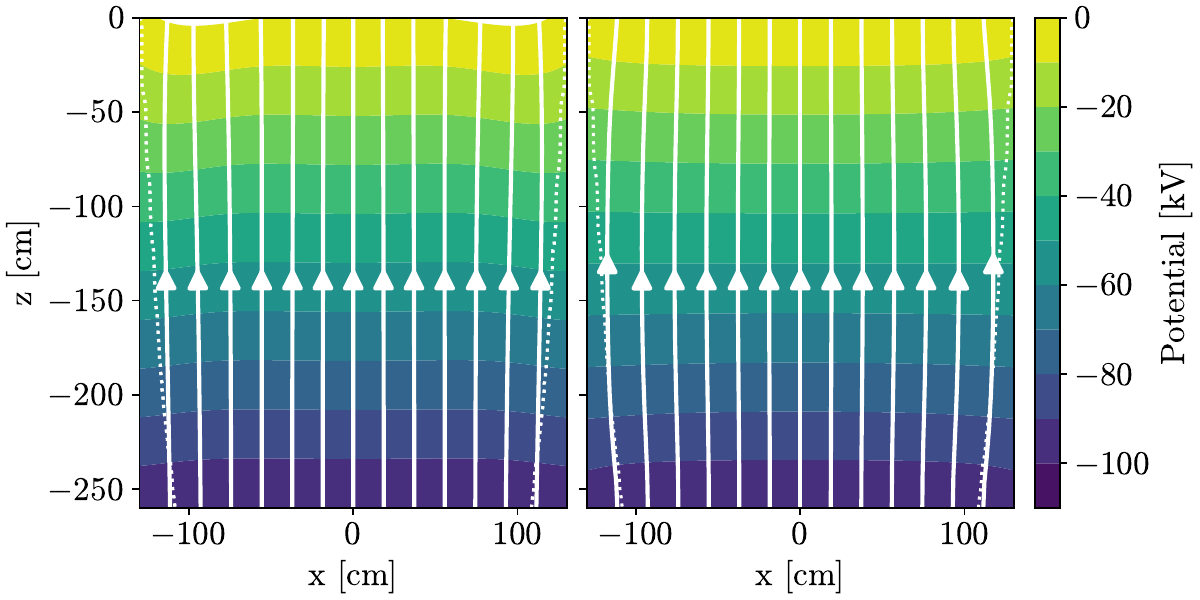}
    \caption{Slices of the learned electric field transformation are displayed in white and overlaid on the scalar potential map at a fixed $y = 0\,$cm as learned by the continuous normalizing flow (left) or derived from the simulation ground truth (right). The edge of the charge insensitive volume, defined by the $p_\mathrm{surv}=0.5$ contour, is denoted by the white dotted lines. The continuous normalizing flow scalar potential map is reconstructed using~\cref{eq:reconstruct_V_from_g}. We can see that the two maps are visually similar.}
    \label{fig:learned-potential}
\end{figure*}

The flow achieves improved position reconstruction accuracy when trained on the same $6 \times 10^5$ events used to generate the FDC map. As shown in~\cref{fig:mse-comparison}, we achieve a similar mean squared error of around 6 cm$^2$ when comparing the corrected positions and the corresponding ground truth positions in a test dataset of $5 \times 10^5$ simulated events. When correcting the same position dataset, a flow trained on $6 \times 10^5$ events performs comparably to an FDC map fitted over $5 \times 10^6$ events, or around one order of magnitude greater in the amount of data. We emphasize that this reduction in necessary calibration data will enable monthly or even weekly electric field monitoring over the course of a science run instead of developing a new FDC map once a year after acquiring enough calibration data.

This improved position reconstruction can be seen when we compare the distributions of the test dataset's event positions before and after corrections from the flow and the FDC map as shown in ~\cref{fig:distribution-compare}, in which the edge of the charge insensitive volume is marked with the white dotted line at $p_{\mathrm{surv}} = 0.5$ and the red solid line marks the edge of the TPC at $r=129.96$ cm. We note that for events near the edge of the charge insensitive volume, both the FDC map and the flow will correct these positions such that some number of events are reconstructed outside of the TPC edge. Unlike the flow's performance, the FDC map's performance is more strongly affected by limited statistics near the edge of the charge insensitive volume, and as such, the FDC is unable to accurately reconstruct event interactions near that edge. On the other hand, the flow corrected distribution is reconstructed closer to the $p_{\mathrm{surv}} = 0.5$ contour at the edge of the charge insensitive volume without as many events extending into the charge insensitive volume. Since we do not constrain the continuous normalizing flow to only produce corrected positions within the TPC, there are still events that are reconstructed outside of the TPC past the extent of the FDC corrected positions.

We further demonstrate the flow's performance by comparing the preservation of local pairwise distances between the corrected position distributions as compared to the ground truth interaction vertex positions in~\cref{fig:bootstrapped-2pcf}. We show the normalized two point correlation function with bootstrapping for both the flow corrected positions and the FDC map corrected positions. These correlation functions are calculated by subtracting the ground truth overdensities. From this, we can see how well each method preserves local pairwise distances between event interaction vertices, and it is evident that the flow preserves more local pairwise distances and clustering for separation distances within one TPC radius. Furthermore, for very local separation distances under 10 cm which is on the order of the binning scheme of the FDC map, the traditional FDC map performs significantly worse than the flow. There is a massive overdensity as seen in~\cref{fig:bootstrapped-2pcf} that is statistically significant, and the flow is able to mitigate this issue significantly. 

Finally, in~\cref{fig:learned-potential} we show that the flow lets us reconstruct the electric field lines and scalar potential map from data. The learned scalar potential and electric field lines can be reconstrud from the flow as following~\cref{eq:reconstruct_V_from_g}. We compare the flow's reconstruction to the ground truth scalar potential and electric field lines derived from the ground truth in simulation from~\cref{sec:simulator-efield-calculation}. Thus, we are able to obtain a physically interpretable and scaled scalar potential map and its consequent negative gradient to acquire the electric field lines. Not only does the flow outperform the histogram-based FDC map when given the same dataset, it also enables us to obtain a physically interpretable scalar potential map from its transformation function and to recover the electric field lines. In a real world context outside of our toy simulation, this means that we can use this to reconstruct the potential field within an experiment.

\section{Conclusions and Discussion}

Ultra rare event searches in noble element TPCs depend on the selection of correctly reconstructed events and the rejection of background events. We have demonstrated that a continuous normalizing flow improves event position reconstruction by modeling the electric field lines inside a noble element TPC and correcting the transverse displacement of ionization electrons. When compared to current developments of FD maps, this physics-informed machine learning model is able to infer an electric field model that is inherently curl-free and differentiable throughout the TPC volume, which is a significant improvement over previous methods of correcting positions from electric field distortions. This method is robust enough to perform well on top array hit patterns from the XLZD toy detector for a realistically problematic electric drift field as based on XENONnT's linear wall charge model. We hypothesize that our method improves data efficiency as it restricts the neural network to produce smooth and conservative representations of the electric field.

Improving position reconstruction will increase the number of interaction events that are correctly reconstructed and result in greater wall background rejection. This is especially critical for a rare event search. For example, the selection of potential WIMP events in a rare event search is limited by the rejection of background events and poorly reconstructed events \cite{XENON:2024xgd}. Furthermore, by applying this continuous normalizing flow to probabilistic position reconstruction, we also quantify the probabilistic uncertainty for the transverse position of an event by propagating the position uncertainties along the learned electric field lines. This enables a fully Bayesian analysis in which the selected interaction events have fully quantified per-event uncertainties and each event's posterior probability distribution is derived. Consequently, it is critical to accurately model these detector effects to ensure the accurate reconstruction of the interaction parameters and to correct for distortions in these parameters. Furthermore, improving the data efficiency of the field distortion correction procedure can enable online monitoring of electric fields and allow for time-dependent corrections.

As next generation noble element TPC experiments come online, the analysis framework must also develop as these experiments continue to scale in size and complexity. With an increased detector volume, larger electrodes, and higher data-taking rates, it is important to develop a robust method for modeling the electric field and subsequently ensure precise interaction event reconstruction. 

\section{Acknowledgments}
We would like to thank Clinton Heider and Erik Engquist for their assistance with the Rice CC*-funded RAPID GPU cluster. The work at Rice University was supported by the Department of Energy AI4HEP program, and the National Science Foundation awards CAREER-PHY-2046549 and OAC-CC* 2019007. SLAC National Accelerator Laboratory is supported by the U.S. Department of Energy, Office of Science under Contract No. DE-AC02-76SF00515.

\bibliography{sn-bibliography}

\appendix

\section{Probabilistic position reconstruction} \label{sec:probabilistic-posrec}

\begin{figure*}[htb]
    \centering
    \includegraphics[width=\linewidth]{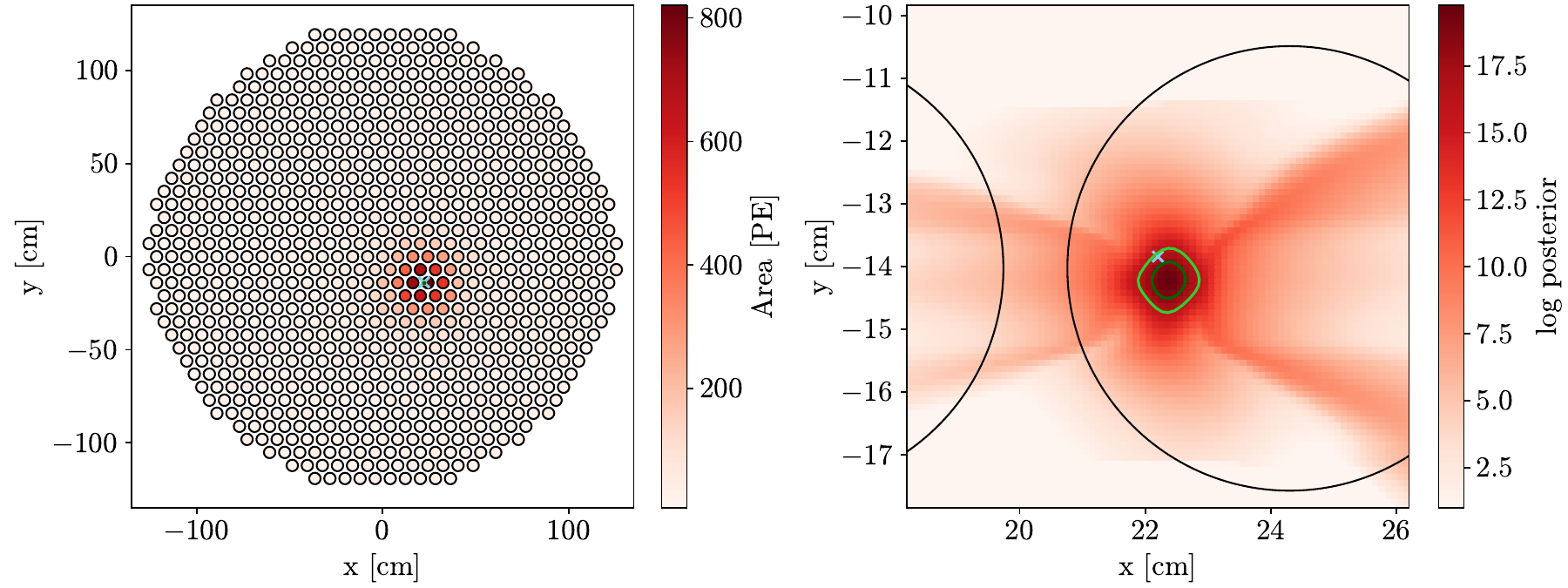}
    \caption{Example of probabilistic extracted position reconstruction given a simulated top photosensor array hit pattern. The true position is marked with a cyan cross, and 68\% and 95\% contours are depicted in dark and light green respectively.}
    \label{fig:posrec}
\end{figure*}

Normalizing flow models are an approach to parameterize probability distributions using neural networks, and are widely used in variational inference~\cite{rezende2015variational, agrawal2020advances} and generative modeling~\cite{esling2019flow, ho2019flow++, zhai2024normalizing}. These models represent probability distributions by describing a transformation on a base transformation, such that 
\begin{equation}
    q_{\boldsymbol{\phi}}(\boldsymbol{\theta}) = q(\boldsymbol{z})\left|\frac{\partial \boldsymbol{f}_{\boldsymbol{\phi}}}{\partial\boldsymbol{z}}\right|^{-1},
\end{equation}
where $q(\boldsymbol{z})$ is the base distribution, $q_{\boldsymbol{\phi}}(\boldsymbol{\theta})$ is the transformed distribution, $\boldsymbol{f}_{\boldsymbol{\phi}} : \mathbb{R}^n \rightarrow \mathbb{R}^n$ is a bijective function constructed with a neural network, and $\boldsymbol{\phi}$ represents the neural network parameters. 

In this section, we will briefly discuss the model we applied for probabilistic position reconstruction, but any position reconstruction model may be used for modeling a TPC electric field following this paper's flow method. A dedicated paper is forthcoming on applying conditional normalizing flows for probabilistically modeling particle positions in physics experiments from photosensor array measurements. It should be noted that position reconstruction is not the primary goal of this work, and indeed the same methods presented here can be used with a more typical position reconstruction method that gives a point prediction.

The position conditioned on the hit pattern is a simulation-based inference problem. We approximate the extracted position's probability density by training a normalizing flow~\cite{JMLR:v22:19-1028} to approximate $p((x_{\mathrm{S2}},y_{\mathrm{S2}})|\mathrm{hit pattern})$, where $(x_{\mathrm{S2}},y_{\mathrm{S2}})$ is the extracted position, and the conditional hit pattern is passed to the neural network. We use a coupling flow with a dense neural network and rational-quadratic splines as the bijective transformation \cite{durkan2019neural}. The transformation architecture includes 6 coupling layers, each containing 128 neural network weights, with rational-quadratic splines parameterized by $\mathrm{K}$=5 knots, and is implemented using \texttt{flowjax}~\cite{ward2023flowjax}.

An example of the probability density of the extracted position given a hit pattern is shown in ~\cref{fig:posrec}.

\section{Heuristic derivation of the change of variables formula for continuous normalizing flows}\label{sec:trace-jac-appendix}

In this section, we derive the change of variables formula~\cref{eq:jac-trace} heuristically; a more rigorous proof can be found in~\cite{NEURIPS2018_69386f6b}. To make this derivation easier to follow, we use the same notation as~\cref{eq:2d_neural_ode_f}. Consider the neural ODE
\begin{equation}
    \frac{\partial \boldsymbol{s}(t)}{\partial t} = \boldsymbol{f'}_{\boldsymbol{\phi}}(\boldsymbol{s}(t), t)
\end{equation}

Given $\boldsymbol{s}(t)$ and an infinitesimal timestep $\delta t$, the transformation $\boldsymbol{h}_{t, t+\delta t}$ represented by solving this neural ODE from time $t$ to $t+\delta t$ is
\begin{equation}\label{eq:h_t_deltat}
\begin{split}
    \boldsymbol{h}_{t, t+\delta t}(\boldsymbol{s}(t)) &= \boldsymbol{s}(t + \delta t)\\ &= \boldsymbol{s}(t) + \boldsymbol{f'}_{\boldsymbol{\phi}}(\boldsymbol{s}(t), t)\delta t.
\end{split}
\end{equation}

Given $p'(\boldsymbol{s})$, following the rules for transforming probability distributions~\cite{Amaral:2024edw}, the change in the transformed log probability density function would be
\begin{equation}
\begin{split}
    \log p_{t+\delta t}(\boldsymbol{h}_{t, t+\delta t}(\boldsymbol{s})) = & \log p_t(\boldsymbol{s}) -\\ 
    & \log\left|\frac{\partial \boldsymbol{h}_{t, t+\delta t}}{\partial \boldsymbol{s}}\right|.
    \end{split}
\end{equation}

Differentiating~\cref{eq:h_t_deltat} gives us a Jacobian determinant of
\begin{equation}\label{eq:jacob-det-h-deltat}
\begin{split}
    \left|\frac{\partial \boldsymbol{h}_{t, t+\delta t}}{\partial \boldsymbol{s}}\right| =& \left|I + \frac{\partial \boldsymbol{f'}_{\boldsymbol{\phi}}}{\partial \boldsymbol{s}}\delta t \right|\\
\end{split}
\end{equation}

Using Levi-Civita symbols $\varepsilon_{i_1 \dots i_n} $, we can expand the determinant of an identity matrix plus a matrix with small coefficients, $|I + \epsilon A|$, as
\begin{equation}\label{eq:identity-plus-epsilon-det}
\begin{split}
    |I + \epsilon A| &= \sum_{i_1,\dots,i_n} \varepsilon_{i_1 \dots i_n} \big((\delta_{1, i_1} + \epsilon a_{1, i_1})\dots\\ 
    &\qquad \qquad (\delta_{n, i_n} + \epsilon a_{n, i_n})\big)\\
    &= \sum_{i_1,\dots,i_n} \varepsilon_{i_1 \dots i_n}  \big(\delta_{1, i_1}\dots \delta_{n, i_n} + \\
    &\qquad \qquad \epsilon a_{1, i_1} \delta_{2, i_2} \delta_{3, i_3}\dots \delta_{n, i_n} + \\
    &\qquad \qquad \delta_{1, i_1} \epsilon a_{2, i_2} \delta_{3, i_3}\dots \delta_{n, i_n} + \\
    &\qquad\qquad\qquad\vdots\\
    &\qquad \qquad \delta_{1, i_1}\dots \delta_{n-1, i_{n-1}}\epsilon a_{n, i_n} +\\
    &\qquad \qquad \mathcal{O}(\epsilon^2)\big)\\
    &= 1 + \epsilon\mathrm{Tr}\left[A\right] + \mathcal{O}(\epsilon^2).
\end{split}
\end{equation}

Using~\cref{eq:identity-plus-epsilon-det} and \cref{eq:jacob-det-h-deltat} and dropping higher order terms in $\delta t$, we can then obtain the change in log probability as
\begin{equation}
\begin{split}
    -\delta \log p &= \log\left|\frac{\partial \boldsymbol{h}_{t, t+\delta t}}{\partial \boldsymbol{s}}\right|\\
    &= \log\left(1 + \mathrm{Tr}\left[\frac{\partial \boldsymbol{f'}_{\boldsymbol{\phi}}}{\partial \boldsymbol{s}}\right]\delta t\right)\\
    &= \mathrm{Tr}\left[\frac{\partial \boldsymbol{f'}_{\boldsymbol{\phi}}}{\partial \boldsymbol{s}}\right]\delta t
\end{split}
\end{equation}

We can thus conclude that
\begin{equation}
    \frac{\partial\log p(\boldsymbol{s}(t))}{\partial t} = -\mathrm{Tr}\left[\frac{\partial \boldsymbol{f'}_{\boldsymbol{\phi}}}{\partial \boldsymbol{s}}\right].
\end{equation}

\section{Continuous normalizing flow training procedure}\label{sec:training-appendix}

The continuous normalizing flow model is structured to have 16 hidden layers, where each hidden layer has a width of 256, implemented in \texttt{JAX}~\cite{jax2018github, deepmind2020jax} and \texttt{equinox}~\cite{kidger2021equinox}. The data input is the two-dimensional position $(x_{\mathrm{S2}}, y_{\mathrm{S2}})$ and the output is one scalar value that can be mapped back to a detector field scalar potential. The position is reconstructed from trained probabilistic position reconstruction model from~\cref{sec:probabilistic-posrec} as conditioned on the event hit pattern. Each data point was integrated through both the extraction field, from time 0 to 10, and the drift field, from time 0 until the time given by the z coordinate associated with its drift time from interaction point to the top of the TPC. The time is given by the $z$-coordinate, essentially scaling by a factor of the drift velocity, and the $z$-coordinate is scaled down by a factor of 5 to improve stability during training. The ODE is solved using the Euler method as implemented in \texttt{diffrax}~\cite{kidger2021on} using a constant step size of 0.5.

The model was trained with a training and validation set, both drawn from one simulated dataset. $6 \times 10^5$ $^{83\mathrm{m}}$Kr events were simulated in a uniform distribution throughout the detector volume. This dataset was cleaned of event positions that were unable to be reconstructed, leaving 533762 hit patterns. From this full dataset, a validation set of 4096 hit patterns is set aside, and the rest of the hit patterns are iterated through for a 65536 training dataset per epoch in a repeated loop. Furthermore, each batch was sharded across 4 NVIDIA A40 GPUs in parallel for training for efficiency. The batch size is 1024 hit patterns with gradient accumulation across every 2 epochs, and when calculating the loss function, 16 samples are generated per hit pattern instance during likelihood estimation. The model shown in this paper was trained using the AdamW optimizer as implemented in \texttt{optax}~\cite{deepmind2020jax} for 40 epochs, with a learning rate scheduler that starts with an initial learning rate of $3\times10^{-4}$, reducing to $1.5\times10^{-4}$ after 20 epochs.

We applied large language models to refactor our model and training code into a clean structured codebase while preserving all functions. We have tested the code to ensure that no logic changes were made during this process.

\end{document}